\documentclass[11pt]{article} 
\usepackage{hyperref} 
\pdfoutput=1

\begin{document} 
 
\title{Droplet Bouncing and Skipping on Microstructured Hydrophobic Surfaces} 
 
\author{Rajat Saksena, Andrew Cannon, Julio Manuel Barros Jr.,\\ William P. King and Kenneth T. Christensen \\ 
\\\vspace{6pt} Mechanical Science and Engineering,\\University of Illinois at Urbana-Champaign.\\ Urbana, IL 61801, USA} 
 
\maketitle

%
 
\textbf{Description}\\ 

The impact of a jet of droplets upon surfaces of varying hydrophobicity is studied via high-speed imaging.  Microstructures on silicone surfaces consisting of cylindrical pillars of varying sizes and spacings are utilized to enhance hydrophobicity.  Comparison of droplet motion after impact with these microstructured surfaces is contrasted with that noted for plain glass (hydrophilic) and flat silicone surfaces.  \href{http://hdl.handle.net/1813/14107}{Fluid dynamics videos} are captured at 6000 fps and played back at 30 fps over a field of view of 1.35\,cm (height)\,$\times$\,2.7\,cm (width) that is back-illuminated with an LED array for 800-micron diameter droplets impinging the surfaces at 2.5 m/s with an angle of incidence of 38$^\circ$ (relative to the surface). Bouncing of droplets after impact is not apparent for the glass and unstructured silicone cases, though many droplets were observed to roll along the surface in the latter case which is consistent with its slightly hydrophobic nature.  In contrast, droplets were found to both skip and bounce upon impacting the microstructured surfaces which indicates a significant enhancement in hydrophobicity due to these surface features.

\end{document}